\documentclass{article}

\usepackage{arxiv}

\usepackage{lineno,hyperref}
\usepackage[normalem]{ulem}
\usepackage{amsmath}
\usepackage{booktabs}
\usepackage[table,xcdraw]{xcolor}
\usepackage{multirow}
\usepackage[numbers]{natbib}

\usepackage[utf8x]{inputenc}
\usepackage{graphicx}

\title{
	Network-Based Approach for Modeling and Analyzing Coronary Angiography
}

\author{
	Babak Ravandi\thanks{Corresponding Author --  \texttt{\url{https://bravandi.net}}}%\thanks{Corresponding author}
	\\
	Department of Computer and Information Technology
	\\
	Purdue University
	\\
	West Lafayette, Indiana 47906, USA
	\\
	\texttt{bravandi@purdue.edu}
	\And
	Arash Ravandi
	\\
	Devision of Orthopeadic Rheumatology
	\\
	Friedrich–Alexander University Erlangen-Nürnberg 
	\\Waldkrankenhaus Erlangen, 91054 Erlangen, Germany
	\\
	\texttt{arash.ravandi@fau.de}
}
\begin{document}

\maketitle

\begin{abstract}

Significant intra-observer and inter-observer variability in the interpretation of coronary angiograms are reported. 
This variability is in part due to the common practices that rely on performing visual inspections by specialists (e.g., the thickness of coronaries). 
Quantitative Coronary Angiography (QCA) approaches are emerging to minimize observer's error and furthermore perform predictions and analysis on angiography images. 
However, QCA approaches suffer from the same problem as they mainly rely on performing visual inspections by utilizing image processing techniques. 

In this work, we propose an approach to model and analyze the entire cardiovascular tree as a complex network derived from coronary angiography images. 
This approach enables to analyze the graph structure of coronary arteries. 
We conduct the assessments of network integration, degree distribution, and controllability on a healthy and a diseased coronary angiogram. 
Through our discussion and assessments, we propose modeling the cardiovascular system as a complex network is an essential phase to fully automate the interpretation of coronary angiographic images. 
We show how network science can provide a new perspective to look at coronary angiograms. 
%Moreover, we discuss how network science provides a new perspective to look at coronary angiograms. 
%Our approach provides a new perspective to look at coronary angiograms. % by utilizing the whole arsenal of network science tools. 

\end{abstract}

%\begin{keyword}
%Complex networks; 
%Cardiovascular system; 
%Angiography;
%%%%Invasive Coronary angiograms; 
%Quantitative coronary angiography;
%%%%Complex systems;
%Coronary network; 
%\end{keyword}

%\linenumbers

%\newpage~\newpage

\section{Introduction}
\label{sec_introduction}

Coronary Heart Disease (CHD) is a major cause of disability and death in developed countries.
%[COPIED] Coronary heart disease (CHD) is a major cause of death and disability in developed countries.
Although over the past four decades CHD mortality rates have declined worldwide, CHD remains responsible for one-third of all deaths in people over age of 35 \cite{uptotdate_epidemiology_coronary_disease}. % \cite{heart_disease_stroke_statistics_2018}. % THIS TAKES WAY MORE SPACE
Invasive coronary angiography is the current gold standard to determine the presence, location, and stage of coronary artery disease as well as to follow-up with the patients after therapeutic procedures \cite{gold_standard_invasive}.
However, potential observer error from performing visual analysis of Coronary Angiograms (CAs) has been estimated to be over 35\% \cite{variability_coronary_arteriograms}.
Quantitative Coronary Angiography (QCA) approaches are emerging to minimize the observer error and further perform  predictions and analysis on angiography images \cite{uptotdate_angiography_clinical}.
Feyter et al. \cite{QCA_limitations_implications_clinical} classified the limitations of QCA approaches to three categories: patient related, technique related, and methodology related.
%Immense technical improvements in imaging techniques (e.g., from 2D to 3D imaging \cite{3D_reconstruction_QCA}) and advances in machine learning have been achieved since Feyter et al. article in 1991.
Subsequently, immense technical improvements in the medical imaging techniques and advances in machine learning have been achieved \cite{retinal_graph_based_approach,MIA_CT_neural_networks,MIA_CT_CNN_classifier};
%,MIA_CT_CNN_classifier
notably, 3D Reconstruction of coronary angiography from 2D images  \cite{3D_reconstruction_QCA,3D_reconstruction_coronary_biplane_angiography}.
%\textbf{3D Reconstruction of Coronary Bifurcation Using Quantitative Coronary Angiography}
However, the main limitation of QCA approaches remains on capturing physiological characteristics such as side branches and bifurcations, hemodynamic assessment, and vasomotion that are technically difficult to measure \cite{QCA_current_era_principles_and_applications,QCA_interventional_cardiology,QCA_physician_determined_computer_generated}.
Hence, several angiography phenomena can lead QCA approaches toward over or underestimation of parameters such as extensive calcium deposits, acute or chronic thrombus, and slow flow \cite{variability_patterns_in_visual_interpret}.
Due to these limitations, QCA approaches lack sufficient accuracy to be employed for clinical purposes \cite{uptotdate_angiography_clinical}.
%We believe the missing key is dynamization of QCA by utilizing the structural characteristics of networks derived from cardiovascular coronary tree.
We believe the missing key is dynamization of QCA; that can be achieved by utilizing the structural characteristics of the cardiovascular coronary tree as a complex network.

%The advantages of utilizing networks arise naturally considering the logic behind network science, i.e., focusing on relations among entities rather than the entities themselves.
The advantage of utilizing networks naturally arises from the way of thinking behind it, that is focusing on the relations among the entities rather than the entities themselves.
For instance, consider the fact that humans and some plants have about 25,000 genes \cite{genes_25000}. 
Having around the same number of genes does not reflect the biological complexity of humans compared to such plants.
Many biologists believe the complexity of an organism arises from the complexity of the interactions between its genes.
The great genome project provided us with the book of life containing the description of all genes, and networks are providing the map of life that describes the dynamics in which genes interact with each other \cite{guided_tour,linked_book}.

\subsection{Innovation}

Heart is a complex system consist of an interconnected network of coronary arteries as the heart's blood supplier.
The innovation of the proposed approach is its ability to create a collective view of the heart's coronary circulation system by capturing the structure of coronary tree.
%The proposed approach enables analyzing network structure-functions relationships, through which, we can identify hidden patterns in coronary networks that relate to \txr{emergent properties$|$[can we use something else?]} such as the existence or formation of stenosis.
Our approach enables analyzing network structure-functions relationships, through which, we can identify hidden patterns in coronary networks. 
Such patterns relate to the formation or existence of conditions such as stenosis. % (calcium buildup) and insufficiency (valve dysfunction or leakage), which usually require repairing or replacing the valves of heart. %https://www.guthrie.org/content/minimally-invasive-heart-surgery-proves-less-more
% Babak --> The following are the main contributions of this work:
%NIMA --> What follows summarizes the major contributions of our work:
The following summarizes contributions of this work: 

\begin{itemize}
	
	%\item We propose a new network-based modeling method to analyze coronary angiography images and connect the structure of cardiovascular system to its functions. 
	
	%\item \txr{We propose a \txr{new} network-based modeling approach that can assist automatizing the analysis of coronary angiography images.}
	
	%\item We propose a network-based modeling approach to assist the automation of interpreting coronary angiography images.
	
	\item We propose a new perspective to analyze and understand coronary angiography images based on capturing the network structure of coronary tree. 
	
%%%%%%%%%%%%
    \item We treat the cardiovascular system as a complex system and present a showcase of three network assessments on a healthy and a diseased coronary angiogram.
	
	%\item We present a showcase of network assessments on a healthy and a diseased coronary angiogram. \sout{to analyze cardiovascular networks.}

%%%%%%%%%%%%

	%\item We view the cardiovascular system as a complex system and discuss how the science of networks is capable of enhancing the existing QCA approaches.
	
	% NIMA VERSION
	%\item We consider a cardiovascular system as a complex network and discuss how the network science can provide insights on the functions of the cardiovascular system given the graph structure of corresponding coronary trees.
	%and connect the structure of cardiovascular system to its functions
	
	% BABAK OLD VERSION
	% \item We view the cardiovascular system as a complex networked system and discuss how network science can provide insights on the functions of cardiovascular system from the structure of coronary trees.
	
	% BABAK NEW VERSION
	%\item We consider the cardiovascular system as a complex system and present three network-based assessments. 
	%Through these assessment, we discuss how network science can provide insights from the graph structure of coronary arteries. 
	
	\item %Through network assessments, 
	We discuss how network science can provide insights from the graph structure of coronary arteries, and ultimately paves the way to fully automate the interpretation of coronary angiography images. % on the functions of cardiovascular system. 
	
\end{itemize}

This article is organized as follows:
%Section \ref{sec_coronary_angiography} provides a comprehensive overview of angiography imaging and QCA achievements and limitations.
%Next, in Section \ref{sec_complex_networks}, we provide the literature on utilizing network science for modeling and analyzing biological systems.
Section \ref{sec_complex_networks} provides a brief overview of employing network science in modeling and understanding biological systems.
In Section \ref{sec_model_case_study}, we introduce our modeling approach by conducting three network-based assessments on two CA. %  and network based assessments by presenting a showcase on two CAs.
%Lastly, we discuss our observations and vision in Section \ref{sec_discussion}.
Lastly, we discuss our vision in Section \ref{sec_discussion}.

\section{Complex Networks and Biological Systems}
\label{sec_complex_networks}

%\txr{This approach, called genetic genomics, integrative
%	genetics, or systems genetics, is proving particularly powerful for the analysis of complex cardiovascular and
%	metabolic traits.}

%The need for quantitative coronary angiography  ...
%This technology utilizes the tools of network science to create a model of heart artery/vein(A/V) network from coronary angiograms.
In recent years, there has been growing interest in complex, self-organizing networks often employed to model the dynamics and structure of complex systems \cite{structure_function_complex_network_newman}.
%TODO add a paper from targeted journal Journal
These are dynamical networks of diffusely interconnected components.
Their behavior is a manifestation of the behavior of the individual components and a reflection of the structural connections between these components.
Examples of complex dynamic graphs abound in nature, from the microscopic cellular level where cells synchronize to perform their functions (heart beating \cite{heart_pacemaker} and neural graphs \cite{measures_of_brain}) to large ecological graphs that respond to perturbations through very slow evolutionary behaviors \cite{transcription_TRN_yeast_2}.
%to understand and control social phenomena such as the political polarization in social networks \cite{polarization}.

%thoroughly review
Lusis and Weiss \cite{cardiovascular_networks} provided a comprehensive review of the advances achieved by employing network science to investigate the cardiovascular system and diseases from the molecular level (genes and proteins). 
%Lusis and Weiss \cite{cardiovascular_networks} reviewed the use of networks in understanding the cardiovascular system and diseases from the molecular level (genes and proteins). 
They showed system-based approaches are likely to play an important role in understanding the  higher-order interactions that lead to formation of diseases such as heart failure, atherosclerosis, cardiac hypertrophy, and arrhythmias.
Moreover, Dashtbozorg et al. \cite{retinal_graph_based_approach} proposed an automated graph-based approach to classify the retinal blood vessels.
Their study was able to label retinal blood vessels with up to 89\% accuracy.
%
%TODO \txr{[find better way to connect two sentences]} 
In another study, Estrada et al. \cite{retinal_topology_estimation} proposed a graph-theoretic framework to classify the retinal blood vessels. 
Their approach obtained an accuracy level up to 93.5\%. 
Furthermore, West et al. \cite{metabolic_scaling_theory_1997} introduced a general model of the circulatory systems as space-filling fractal networks. 
Their model derives the well known biological scaling relationship (i.e., \textit{metabolic-rate} $\propto$ \textit{body-mass}$^{3/4}$) shedding light on the evolution of biological systems. 
%The above studies demonstrate the practicality and advantages of modeling blood vessels as a complex network and utilizing the network sciences for analyzing them.
The above studies demonstrate the practicality and advantages of modeling blood vessels as a complex network and utilizing network sciences to analyze the graph structure of the circulatory system.
%The above studies demonstrate the practicality and advantages of modeling blood vessels as a complex network and utilizing the network sciences to analyze body organs. 

%\pagebreak

\section{Proposed Approach and Case Study}
\label{sec_model_case_study}

\begin{figure}[t!]
	\centering
	%\includegraphics[width=\columnwidth]{fig1_creation_process.pdf}
	%0.70
	\includegraphics[scale=0.62]{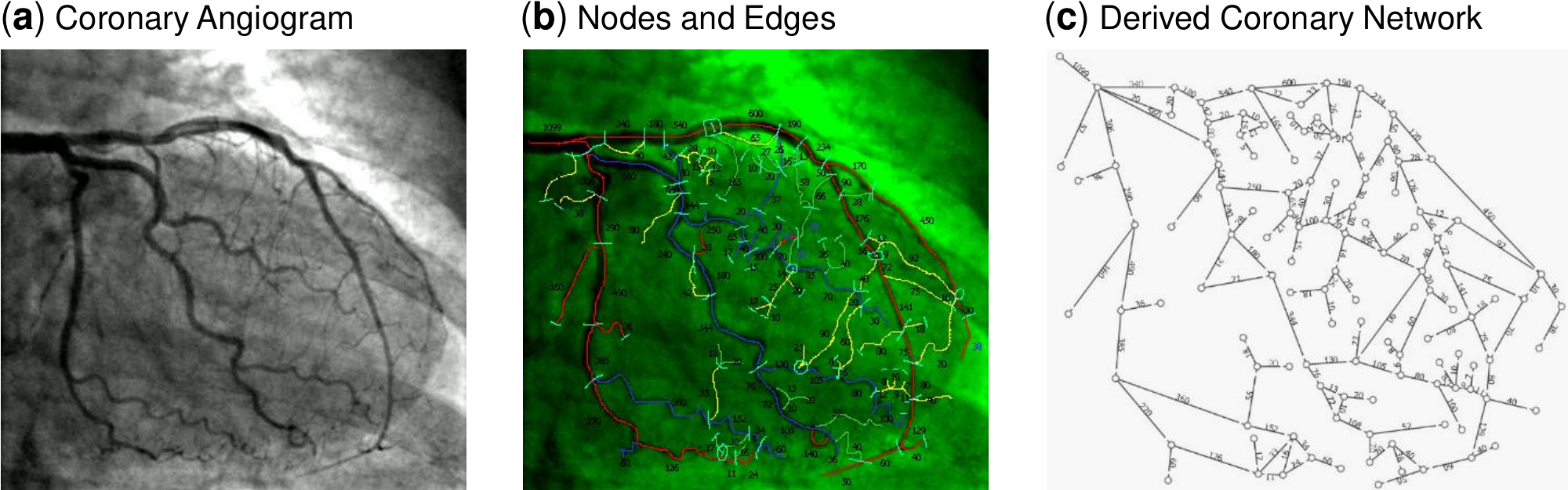}
	\caption{
		An example of network creation process%(\txr{Nima: caption needs more details!})
	}
	\label{fig_network_creation}
\end{figure}

In this section, we propose our model by presenting a case study for both healthy and diseased CAs. 
The case study concentrates only on the Left Coronary Arteries (LCA).
We label an angiogram as diseased if a stenosis exists in the LCA. 
However, without loss of generality, the proposed model is naturally extendable to integrate all cardiac vessels and provide a complete map of heart coronary arteries. 
Figure \ref{fig_network_creation} illustrates a CA and the process to derive a network of coronary vessels. 
%TODO decide to use 'edge' or 'link'
%The intersections of vessels represent the network nodes, and the weighted edges represent vessels with their diameter captured by the weight of edges. 
A network consists of a set of nodes (representing a system entities) and a set of edges (capturing a relationship between those entities). 
In the proposed model, a node represents an intersection between vessels, and a weighted edge represents a vessel.
The weight of an edge is calculated by multiplying the diameter of a vessel by its length. 
%The process of creating coronary networks starts with: a) Identifying the intersections of vessels (i.e., nodes), and b) measuring the length and diameter of each sub-vessels between the identified intersections to calculate the weights of edges. 
Two steps were taken to create the coronary networks in this work: 1) identify the intersections of vessels (i.e., nodes), and 
2) measure the length and diameter of each sub-vessel between the identified nodes and calculate the weights of edges. 
We employed graphical filters to magnify the vessels as presented in Fig. \ref{fig_network_creation} (b) and manually conducted these steps. 
However, without loss of generality, one can fully automate the network creation process by employing the variety of tools developed for performing visual inspections on angiography images 
\cite{QCA_hybrid_coronary_bifurcations,3D_reconstruction_QCA,computer_assisted_diagnosis_coronary,QCA_current_era_principles_and_applications}. 
Figure \ref{fig_network_creation} (c) presents the created weighted coronary network.

\subsection{Healthy and Diseased Coronary Networks}

\begin{figure}[b!]
	\centering
	%scale=0.55
	\includegraphics[scale=0.55]{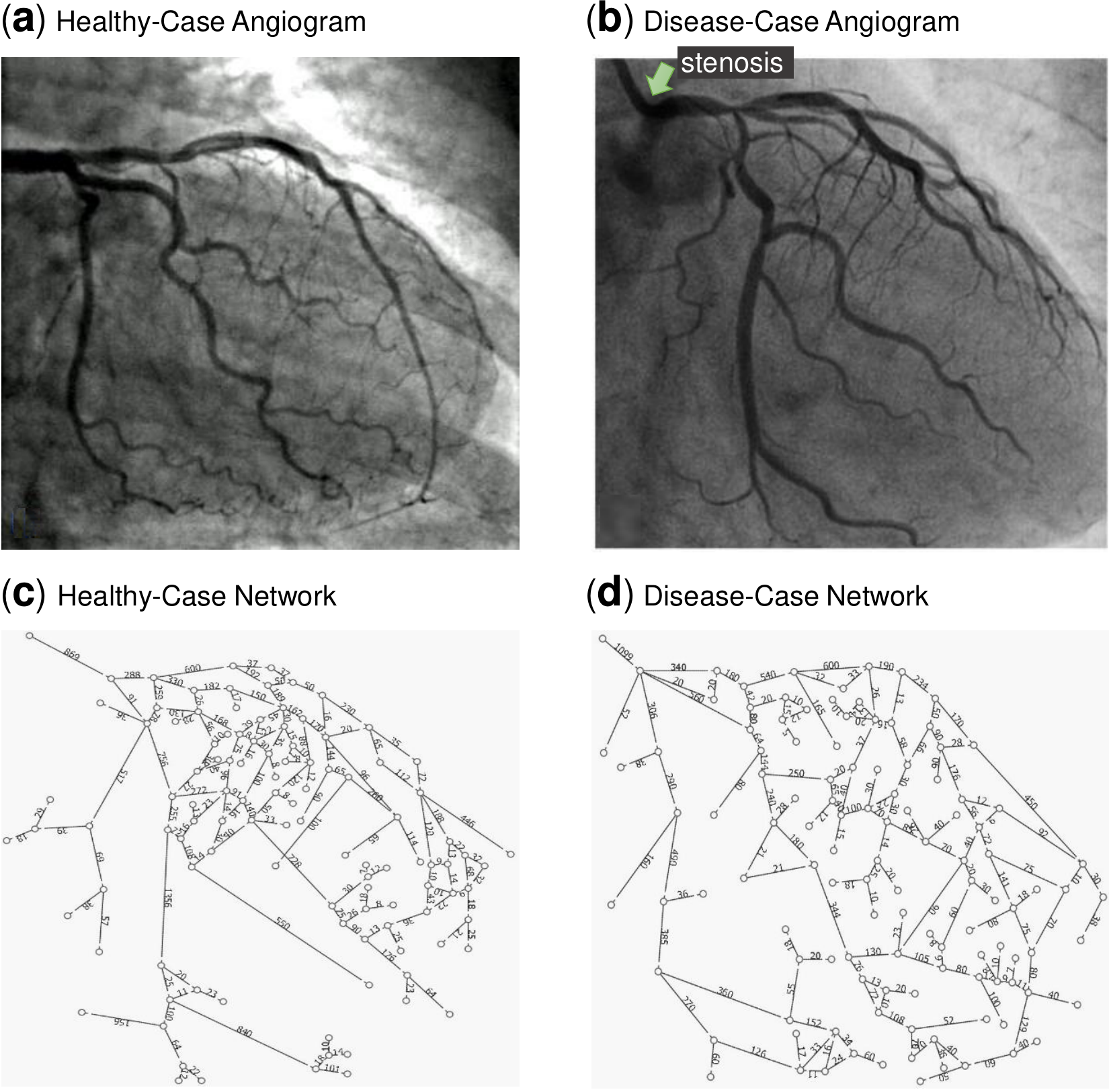}
	\caption{
		%Examples of healthy and disease heart’s coronary angiogram with their corresponding networks.
		%Healthy and disease case coronary angiograms with their corresponding networks.
		Healthy and disease-case coronary angiograms and their networks.
	}
	\label{fig_angiograms}
\end{figure}

Figure \ref{fig_angiograms} presents angiograms for both a healthy and a diseased heart alongside their corresponding coronary networks.
The healthy angiogram is collected from \cite{healthy_case_angiogram} and the source of diseased angiogram is in \cite{disease_case_angiogram}. 
In the diseased angiogram, a stenosis is marked by the green arrow. 
%Left coronary artery
%Our main purpose by introducing the angiograms in Fig. \ref{fig_angiograms} is to use them as a sample to introduce our approach (the healthy-case and disease-case angiograms are not related to each other). 
The healthy-case and disease-case angiograms in Fig. \ref{fig_angiograms} are not related to each other. 
Our goal is to utilize the CAs in Fig. \ref{fig_angiograms} to introduce our modeling approach. 
The global network characteristics \cite{structure_function_complex_network_newman} of healthy-case and disease-case networks are summarized in Table \ref{tbl_network_characteristics}, in the following five columns:
1) \textit{number of nodes} represents the number of intersections between the vessels,
2) \textit{number of edges} represents the number of vessels,
3) \textit{average degree} presents the average number of connections of the nodes,
%(3) the average degree denoted by $\langle k\rangle$, presents the average number of connections of the nodes,
%4) \textit{clustering coefficient} shows the number of triangles in the network, and
4) \textit{average clustering coefficient} captures the degree of connectedness among neighbors of nodes, and 
%NetworkSience book --> degree to which the neighbors of a given node link to each other
%%%
%(5) \textit{diameter} of a network presents the length of the longest path in the combination of all shortest paths between every two nodes.
5) \textit{diameter length} of a network presents the length of the longest shortest path between all combinations of nodes.
At the first glance, the average clustering coefficient of the disease-case network is relatively smaller than the healthy-case by 36\%.

\begin{table}[t!]
	\centering
	\caption{
		\label{tbl_network_characteristics}
		%Global network characteristics of the angiograms
		%Global network characteristics of coronary networks 
		Coronary network characteristics
	}
	\begin{tabular}{p{7em}ccccc}
		\toprule
		%\hline
		Network & \multicolumn{1}{p{4em}}{Number of nodes} & \multicolumn{1}{p{4.0em}}{Number of edges} & \multicolumn{1}{p{4.0em}}{Average degree %$\langle k\rangle$
		} &  \multicolumn{1}{p{4.7em}}{Average clustering  coefficient 
		} & \multicolumn{1}{p{4.0em}}{Diameter length} \\
		\midrule
		%\hline
		Healthy-Case & 115   & 140   & 2.4348 & 0.099 & 23 \\
		\midrule
		%\hline
		Disease-Case & 109   & 138   & 2.5321 & 0.063 & 24 \\
		%\hline
		\bottomrule
	\end{tabular}%
\end{table}%

%\clearpage

\subsection{Network Visualization}

\begin{figure}[b!]
	\centering
	%scale=0.555
	%\includegraphics[scale=0.45]{fig3_visualization.pdf}
	\includegraphics[width=\columnwidth]{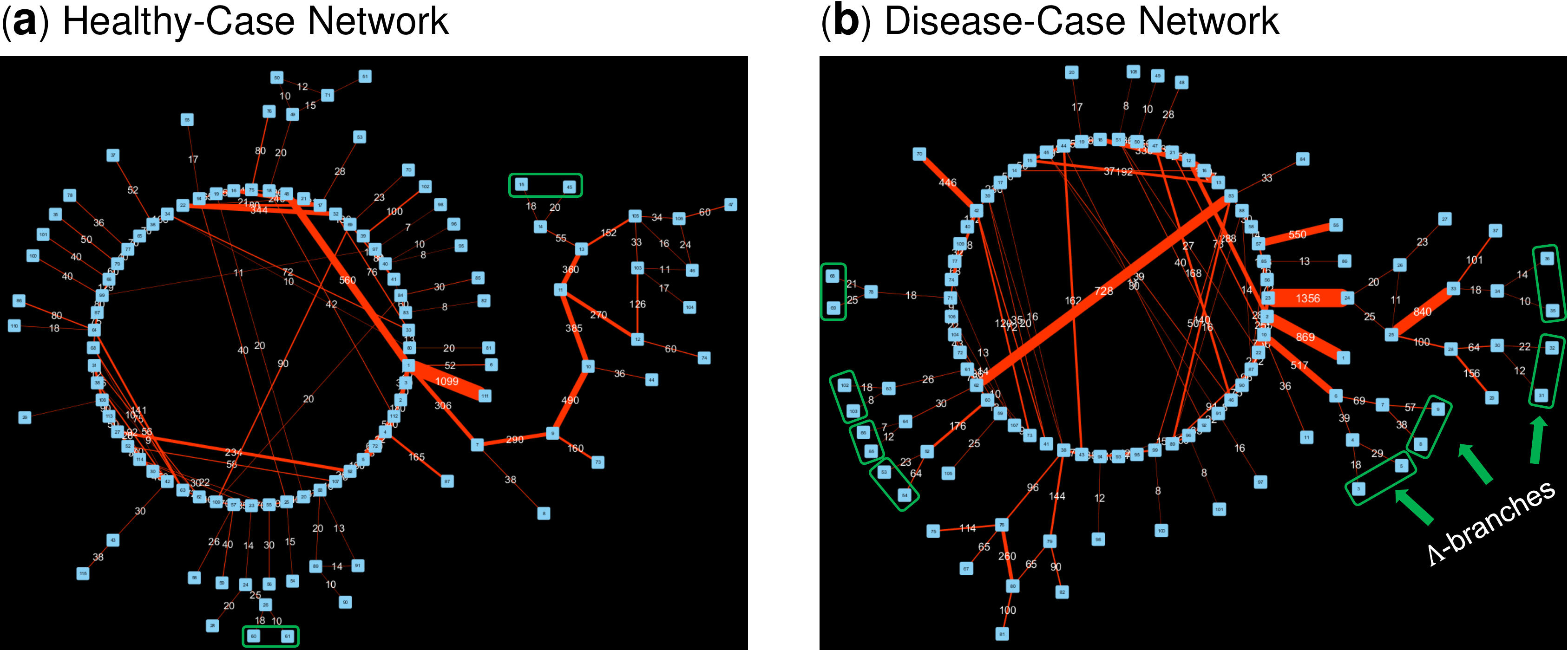}
	\caption{
		Coronary networks visualizations.
		Thickness of edges indicate the vessel's diameter times their length and the green boxes mark $\Lambda$-branches.
		%The disease-case network has several more $\Lambda$-branches compared to the healthy-case network.
	}
	\label{fig_visualization}
\end{figure}

\begin{figure}[b!]
	\centering
	%\includegraphics[scale=0.65]{fig3.pdf}
	%0.75
	\includegraphics[scale=0.70,page=1]{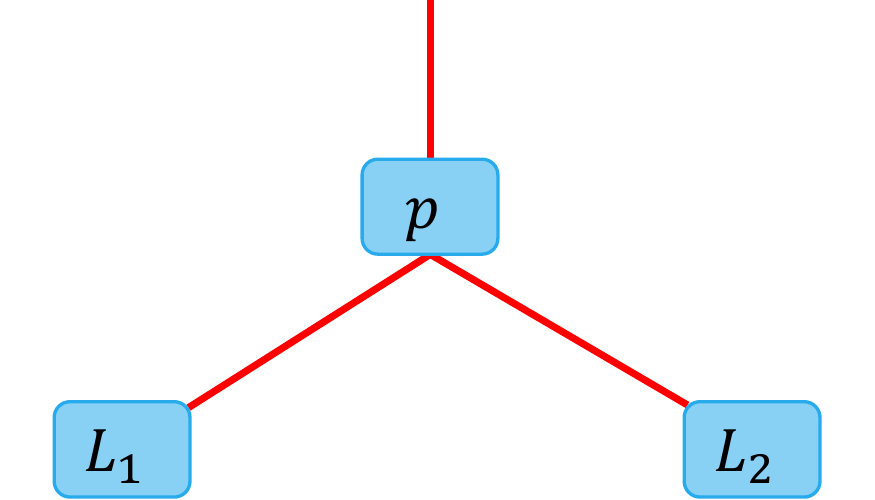}
	\caption{
		%$\Lambda$-shape-leaf structure.
		$\Lambda$-branch structure.
	}
	\label{fig_v_shape_branch}
\end{figure}

%By visualizing networks, we can get insights on their structure and patterns of connections.
Visualizations of networks may provide insights on their structure and patterns of connections.
Figure \ref{fig_visualization} illustrates the derived coronary networks in a circular layout. 
The thickness of edges (i.e., weights) represents the diameter of vessels multiplied by their length (the unit of measurements is pixel). 
Also, the green boxes in Fig. \ref{fig_visualization} mark $\Lambda$-branches as illustrated by Fig. \ref{fig_v_shape_branch}.
%A $\Lambda$-branch consists of one parent node with only has two children that are not connected to any other node (i.e., tree leaves).
A $\Lambda$-branch consists of a single parent node that only has two children (i.e., the coronary tree leaves) who are not connected to any other nodes. 
Also, the parent node must only have a single additional connection other than its children.


The disease-case network has several more $\Lambda$-branches compared to the healthy-case network. 
%This could indicate that in the disease heart blood is not being properly supplied.
This indicates blood is not being properly supplied to the diseased heart.
The abundance of $\Lambda$-branches could reflect the Neovascularization phenomenon \cite{neovascularization}, which happens when the blood is not being properly supplied and the heart starts creating new vessels.
%These bridges can be observed in Fig. \ref{fig_angiograms} (b) where we can find small vessels are connecting the main arteries. 
These vessels can be observed in Fig. \ref{fig_angiograms} (b) where many small vessels are emerged from the main arteries. 
%In the next section, by analyzing the degree distribution of coronary networks, we show how to capture this behavior by only using the degree distributions of the network.
In the next section, we show how to systematically capture this behavior by analyzing the degree distribution of coronary networks.

\subsection{Assessment of Degree Distribution}

\begin{figure}[b!]
	\centering
	%\includegraphics[scale=0.65]{fig3.pdf}
	%0.5
	\includegraphics[scale=0.56,page=2]{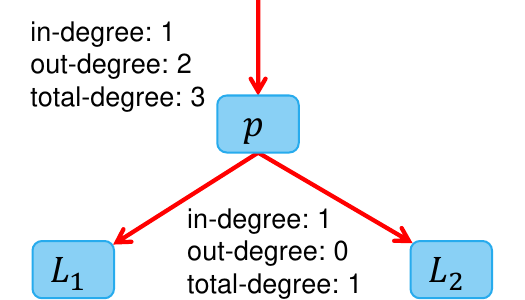}
	\caption{
		Directed $\Lambda$-branch structure.% with in/out/total-degrees of its nodes.
	}
	\label{fig_v_shape_branch_directed}
\end{figure}
\begin{figure}[b!]
	\centering
	%\includegraphics[width=\columnwidth]{fig6_deg_dist.pdf}
	%0.68
	\includegraphics[scale=0.57]{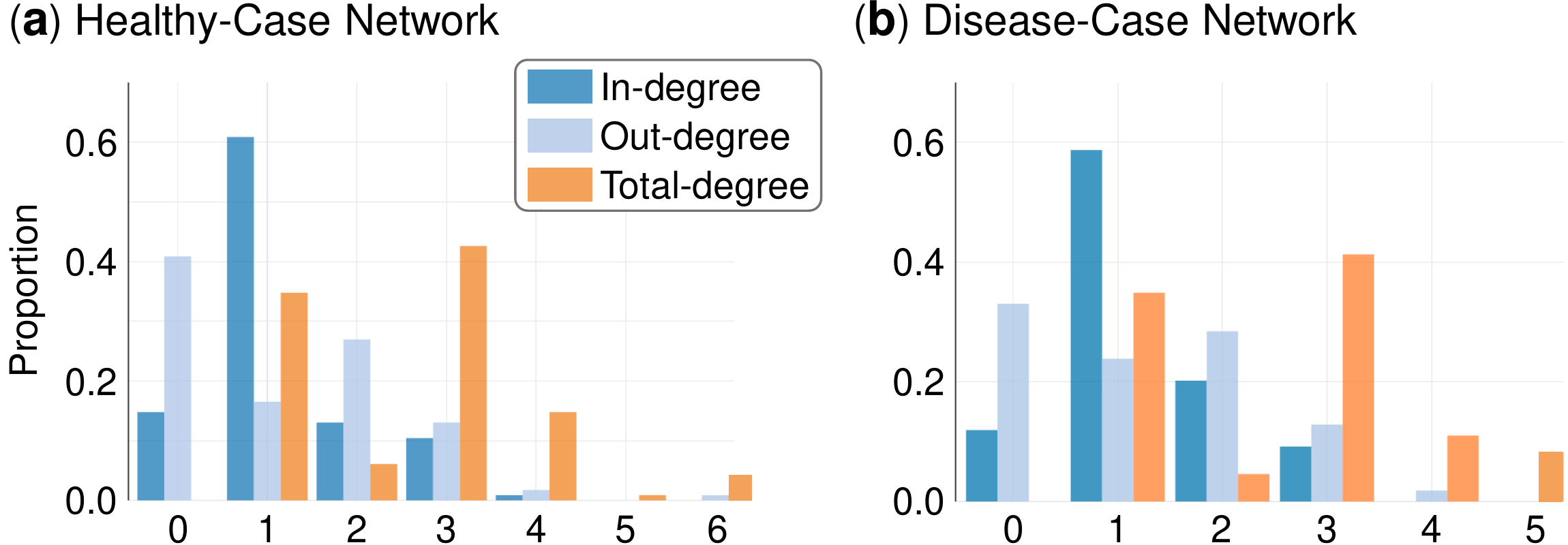}
	\caption{
		Degree distributions of coronary networks.
	}
	\label{fig_degree_dist}
\end{figure}
\begin{figure}[b!]
	\centering
	%\includegraphics[width=\columnwidth]{fig7_quartiles.pdf}
	%0.46
	\includegraphics[scale=0.54]{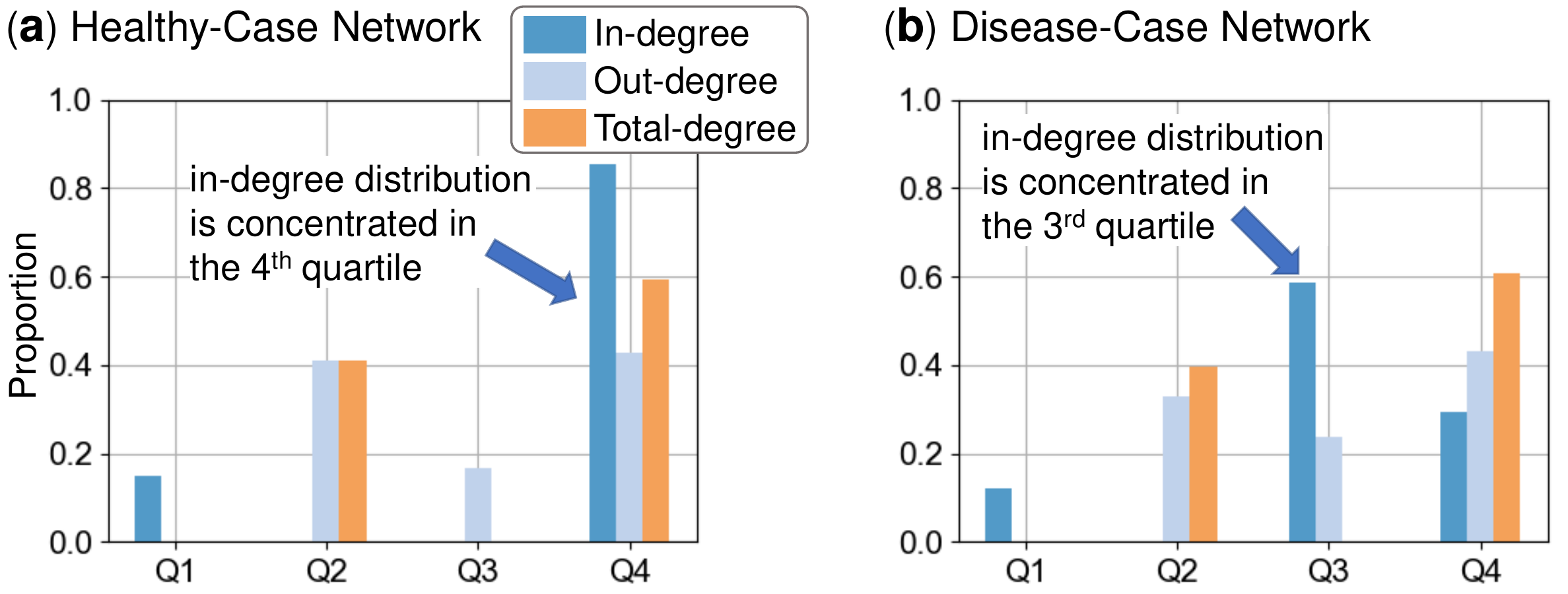}
	\caption{
		Quartile degree distributions of coronary networks.
	}
	\label{fig_quartile_deg_dist}
\end{figure}

The degree distribution of a network represents the distribution of connections among nodes. 
In the coronary networks, the degree distribution presents the extent, in which, vessels are connected to each other.
Blood flows in a fixed direction in human's cardiovascular system.
Hence, we employ directed edges to capture the direction of blood flow. 
Figure \ref{fig_v_shape_branch_directed} illustrates a directed $\Lambda$-branch with the degrees of its nodes. 
For a give node, the \textit{total-degree} indicates its number of connections, the \textit{in-degree} indicates the number of connections to the node, and the \textit{out-degree} indicates the number of connections from the node.

Figure \ref{fig_degree_dist} presents the in-degree, out-degree, and total-degree distributions of the healthy-case and disease-case networks.
At the first glance, there is no significant difference between the degree distributions of the coronary networks.
However, a significant difference is observed by comparing the quartile-degree distributions of the healthy-case and disease-case networks, which is presented in Fig. \ref{fig_quartile_deg_dist}. 
In the healthy-case network, most nodes are concentrated in the fourth quartile for all three degree distributions.
However, in the disease-case network, the concentration of in-degree distribution is shifted to the third quartile.
%TODO 'This behavior' do not call it behavior
%This behavior arises from the existence of many directed $\Lambda$-branches in the disease network.
This shift is due to the abundance of directed $\Lambda$-branches in the disease-case network.
%To conclude, the analysis of degree-distribution could be used to determine the extent, in which, a heart tries to create new vessels. % (the neovascularization phenomenon).
To conclude, the patterns of connections in coronary networks could provide insights on the condition of the cardio vascular system. 
For example, the analysis of degree-distribution could be used to determine the extent, in which, a heart is trying to create new vessels to overcome inefficient blood circulation. % (the neovascularization phenomenon).

\subsection{Assessment of Network Integration}

\begin{figure}[b!]
	\centering
	%0.263
	%\includegraphics[scale=0.35]{fig8_integration.pdf}
	\includegraphics[width=\columnwidth]{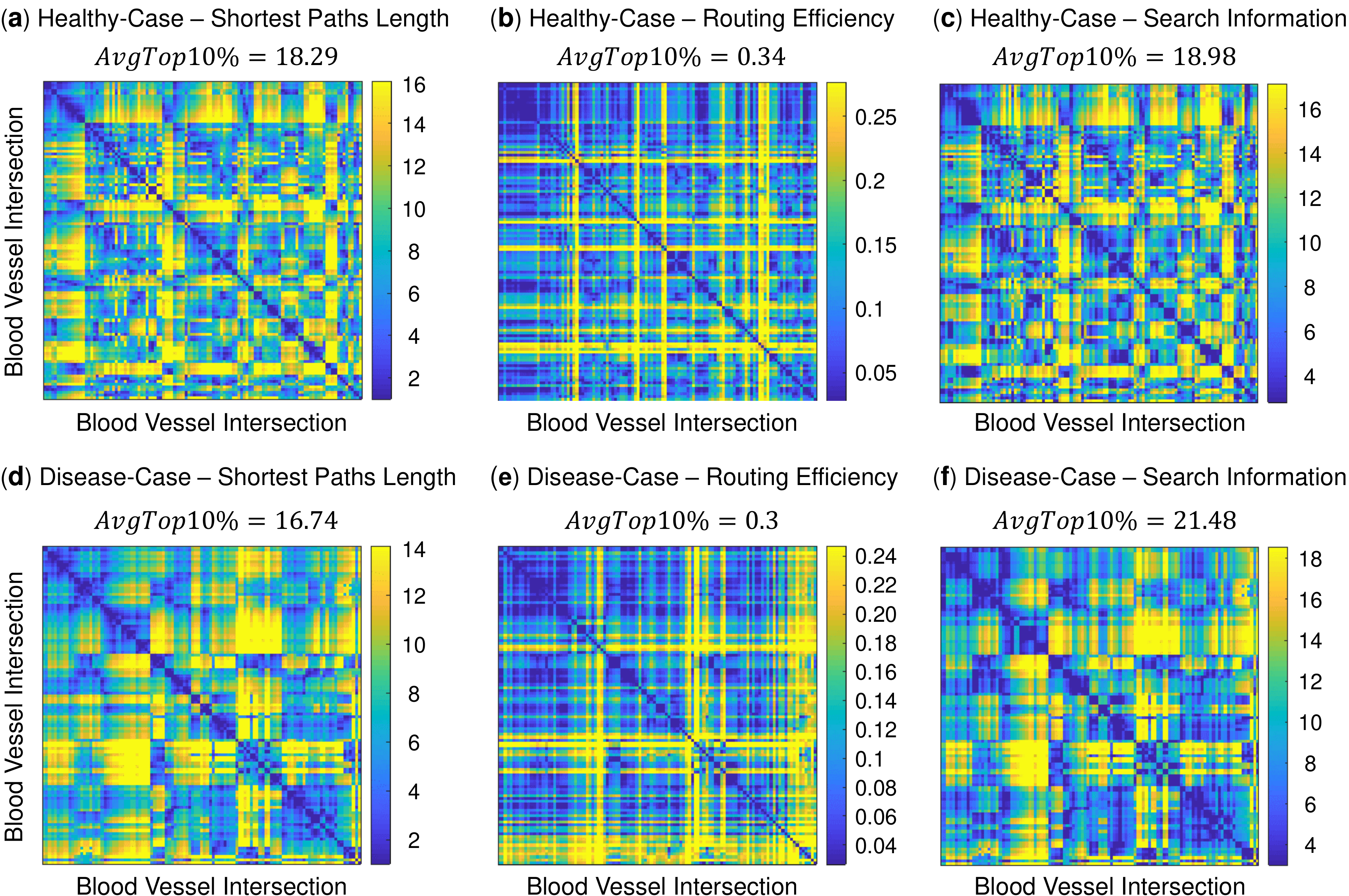}
	\caption{
		Analysis of network integration between all pairs of nodes.
	}
	\label{fig_integration_analysis}
\end{figure}

The efficiency of a network is the measurement of how efficiently it exchanges information.
In transportation networks, this measurement corresponds to the efficiency of patrons commuting in terms of time and distance.
We can utilize the patterns of connections in structure of systems to infer their functional efficiency \cite{efficiency_routing}.
%We can utilize the underlying general principles of a system constructions to infer their functional efficiency \cite{efficiency_routing}.
%The assessment of integration in the coronary networks corresponds to measuring the efficiency of blood circulation in the cardiovascular system.
The assessment of integration in the coronary networks corresponds to quantifying the efficiency of blood circulation in the cardiovascular system.

Figure \ref{fig_integration_analysis} provides three measures of network integration: \textit{shortest-paths length}, \textit{routing-efficiency}, and \textit{search-information}.
The \textit{shortest-paths length} provides the least number of hops (i.e., edges) that needs to be taken to navigate from any source node to any destination node \cite{structure_function_complex_network_newman}. % \cite{integration_communication_boundaries}. %% NO SPACE TO ADD A NEW REF
Figure \ref{fig_integration_analysis} (a) and (d) present the lengths of shortest-paths  between all pairs of nodes.

The \textit{routing efficiency}, also known as global efficiency enables to quantify how cost-efficient a particular network is, where the cost depends on the weight of edges  \cite{communication_efficiency_morphospace}.
Hence, this assessment enables to include the vessel's diameter and length (i.e., weight of edges) in quantifying the efficiency of blood circulation. 
For all pairs of nodes, we present this measurement in Fig. \ref{fig_integration_analysis} (b) and (e).

Lastly, the \textit{search-information} quantifies the amount of information needed for a walker to perform an efficient routing (i.e., quantify accessibility or hiddenness).
That is, how much information is needed for a walker to walk on a shortest path when the walker randomly travels between the nodes \cite{integration_hide_and_seek,integration_searchability}. %brain_communication_analytic
Figure \ref{fig_integration_analysis} (c) and (f) present this measurement between all pairs of nodes in the healthy-case and disease-case coronary networks. 

Through the assessment of network integration, we observe that the healthy-case network requires less information to find efficient routes.
%The main outcome from the assessment of network integration is that the healthy network requires less information to perform an efficient routing.
%In other words, the \textit{search-information} measurement of healthy network is smaller (less hidden) compared to the disease-case network.
In other words, shortest paths are less hidden in the healthy-case network compared to the disease-case (i.e., smaller \textit{search-information}). 
This observation indicates the measurement of \textit{search-information} could be used as a feature to classify healthy and diseased coronary networks.

\subsection{Assessment of Controllability}

\begin{figure}[t!]
	\centering
	%0.42
	%\includegraphics[scale=0.63,page=2]{fig9_controllability.pdf}
	\includegraphics[width=\textwidth,page=1]{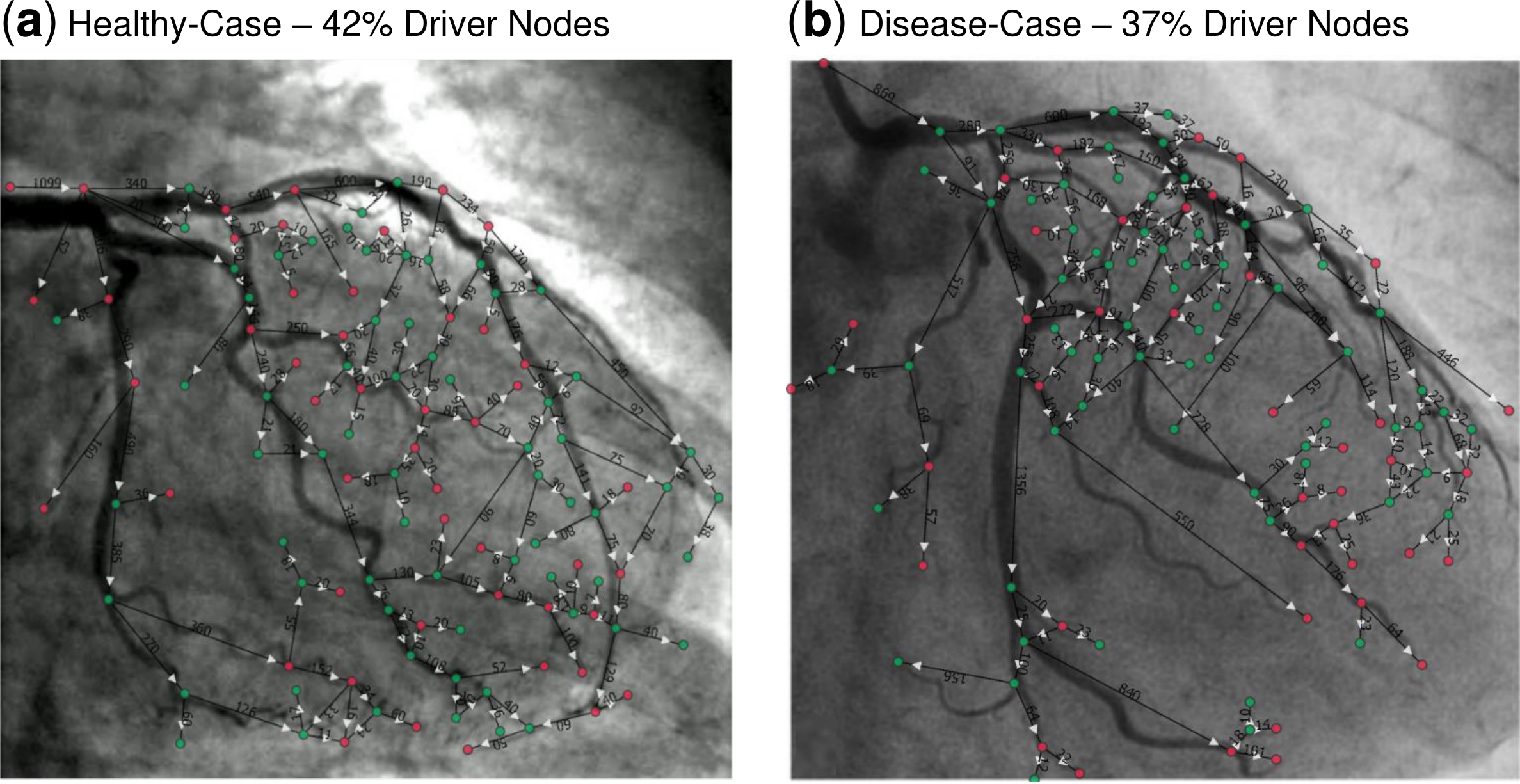}
	\caption{
		The controllability assessment (driver nodes are marked with red color). 
		%Controllability of the coronary networks (red nodes are drivers).
		%Driver nodes are marked with color red.
	}
	\label{fig_controllability}
\end{figure}

The controllability of complex networks is the study of controlling the state of networks from any initial value to a desired value in finite time via stimulating a set of key nodes called \textit{driver nodes}.
Efficient algorithms are developed to identify driver nodes in complex networks \cite{controllability_complex_networks,heuristic_temporal_network}.
%\cite{matrices_matroids_control_max_matching}
Most control scenarios are interested in identifying a minimum number of driver nodes needed to control a system. 
In coronary networks, this is analogous with controlling the flow of blood by modifying the flow that can pass through each node (arteries' intersections).
Figure \ref{fig_controllability} presents the driver nodes (marked red) for both coronary networks. % healthy and disease-case networks.

Intuitively, being easy to control (for cardiovascular systems) might be taken as a sign for healthiness. 
However, having a small percentage of driver nodes in a coronary network indicates a small number of malfunctions can perturb the whole system. 
Hence, a healthy network with a high percentage of driver nodes is more resilient to malfunctions. % (e.g., Cardiac ischemia--lack of blood flow and oxygen to the heart muscle). % blood not being properly supplied to all areas).
Figure \ref{fig_controllability} shows the disease-case network has less driver nodes (37\%) compared to the healthy-case network (42\%). 
%In general, this means the disease-case network is easier to control \txr{[Maybe remove this sentence]}. 

%\clearpage

%\clearpage
%\pagebreak

\section{Discussion and Conclusion}
\label{sec_discussion}

%\txg{
%[GOOD FOR DISCUSSION]
%In addition : it remains a bidimensional representation of a tridimensional complex structure, which can represent a source of error in measurements \cite{QCA_limits_intermediate_stenosis_measuring}. \txr{[DOES NOT WORTH TO CITE|WERE you able to download this article?]}
%}

\begin{figure}[t!]
	\centering
	\includegraphics[width=\textwidth]{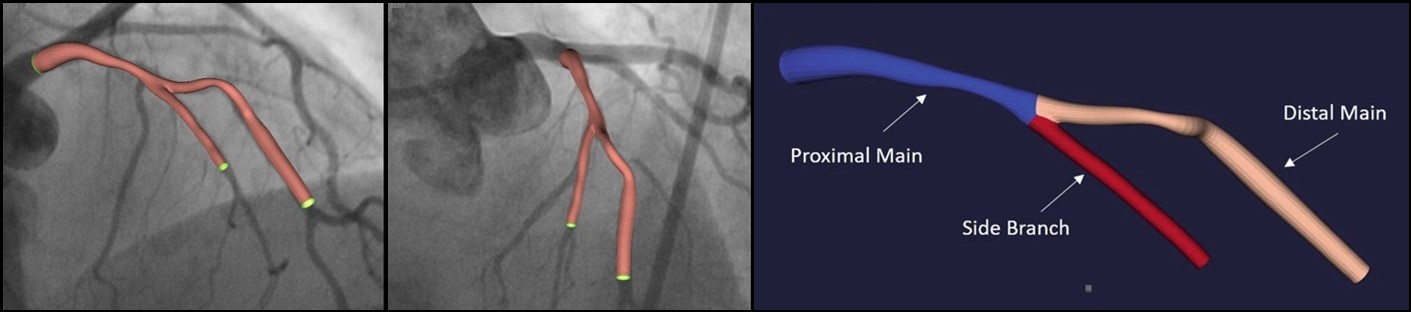}
	\caption{
		3D reconstruction of coronary angiograms (courtesy Andrikos et al. \cite{3D_reconstruction_QCA}).
	}
	\label{fig_3D_reconstruction}
\end{figure}

The predominant methods to identify cardiovascular conditions primarily focus on analyzing the visual properties of coronary arteries (e.g., the diameter of arteries).
For instance, Soroushmehr et al. \cite{computer_assisted_diagnosis_coronary} proposed a QCA approach to assist the diagnosis of CAs. 
%To the best of our knowledge, Soroushmehr et al. proposed the closest existing work in automating the detection of stenosis based on the heart coronary angiograms is \cite{computer_assisted_diagnosis_coronary} with the title ``Computer-assisted Diagnosis of Coronary Angiography'' (the abstract is attached to this document).
%Their approach is primarily based on the visual properties of coronary arteries (e.g., thickness of the arteries) and it can be extensively improved by our proposed approach. 
Their approach is primarily based on the visual properties of coronary arteries (e.g., thickness of the arteries) and it can be extended by employing network science. 
%it can be combined with the proposed network-based approach to enhance the accuracy and extend the functionality. %fully automate the diagnosis of CAs. 
In addition to employing the visual properties of CAs, our proposed approach enables to analyze the dynamics of cardiovascular system. 
Moreover, Andrikos et al.  \cite{3D_reconstruction_QCA} introduced a novel approach for 3D reconstruction of CAs as illustrated in 
Fig. \ref{fig_3D_reconstruction}. 
Their approach can be naturally utilized to automate the process of network construction from CAs. % and incorporate our proposed model. 

%The proposed model provides the basis for development of a new systematic methodology to study and diagnose the cardiovascular system based on coronary angiograms.
The proposed modeling approach provides the basis for development of a new systematic methodology to study the cardiovascular system and automate the diagnosis of  coronary network pathology.
An advantage of such a methodology is introducing new features based on network measurements such as the routing efficiency and controllability. 
For instance, these features could be used for the early detection of cardiovascular pathology by training machine learning classifiers and developing network-based diagnostic methods. 
%%Also, the existing approaches may improve their diagnostic methods by utilizing network-based features. % and consequently improve the chance of early detection. 
%Hence, by employing the aforementioned features, the proposed modeling approach may improve the existing diagnostic methods by increasing the accuracy of learning models and consequently improve the chance of early detection.
%These models can be used to enable a much better therapeutic prognosis since having more features could increase the accuracy of learning models and consequently increase the chance of early detection of cardiovascular conditions.
%Hence, it enables a much better therapeutic prognosis since having more features increases the accuracy of learning models and consequently the chance of early detection of cardiovascular conditions.
Similarly, the proposed approach can improve the accuracy of procedure follow-ups such as the early detection of revascularization after stent implantation.
%Another important advantage of having a systematic methodology capable of automating the interpretation of CAs is minimizing human error.
Another important advantage of developing a systematic methodology is minimizing human error that accounts for a significant observer error \cite{variability_coronary_arteriograms}.

%We should mention that our theory not only can help to automatizing invasive and noninvasive angiography and also more detailed complementary data bridging to achieve the better outcome.

Furthermore, non-invasive coronary angiography such as Computed Tomography Angiography (CTA) are already of significant value in the diagnostic procedure of patients. %with low-to-moderate pretest probability. 
Our modeling approach can enhance the current literature on computer-based approaches for the interpretation of CTA images \cite{MIA_CT_CNN_classifier,MIA_CT_neural_networks}. %by utilizing network-based measurements . 
%Furthermore, the proposed approach could be used to investigate the possibility of surpassing the limitations of current computer-based approaches that interpret coronary angiography images such as QCA approaches.

The proposed network-based approach paves the way to apply the whole arsenal of network science tools on analyzing and classifying the CAs. 
%However, the authors acknowledge future work is necessary to conduct rigorous studies with more than two CAs and further formalize and validate this approach. 
%However, future work is necessary to conduct rigorous studies with more than two CAs and further formalize and validate this approach.
However, the authors acknowledge that a 	rigorous study with more than two CAs should be done to further formalize and validate this approach. %the proposed approach in this paper. 
%Such a study can be conducted as a future work. % in this direction. 

\section{Acknowledgements}

The authors acknowledge Professor Joaqu{\'\i}n Go{\~n}i, School of Industrial Engineering at Purdue University, West Lafayette, USA and Dr. Sophoclis Sophocleous, Pulmonology Resident in Bethanien Hospital, Solingen, Germany for their help and guidance on this paper. 
We like to thank Mr. Javad Darivandpour, Ph.D. candidate in the Department of Computer Science at Purdue University, West Lafayette, USA for his constructive criticism of the manuscript.

%\section*{Acknowledgements}

%The authors acknowledge Dr. Joaquín Goñi for his help and guidance on this article.

%\section*{References}

\bibliographystyle{unsrt}  
\bibliography{references}

\end{document}